\documentclass[aps,prc,superscriptaddress,twoside,twocolumn,nofootinbib,10pt,%
showpacs,floatfix]{revtex4-1}

\usepackage{amsmath,amssymb}
\usepackage{graphicx,bm}
\usepackage{epstopdf}
\usepackage{ulem}
\usepackage[usenames]{color}
\usepackage{float}
\usepackage{color}
\newcommand{\Slash}[1]{{\ooalign{\hfil/\hfil\crcr$#1$}}}

\allowdisplaybreaks

\begin{document}

\title{
Relation between the
mass modification of the heavy-light mesons and the chiral symmetry structure in dense matter
}

\author{Masayasu Harada}
\email{harada@hken.phys.nagoya-u.ac.jp}
\affiliation{Department of Physics,  Nagoya University, Nagoya, 464-8602, Japan}

\author{Yong-Liang Ma}
\email{yongliangma@jlu.edu.cn}
\affiliation{Center for Theoretical Physics and College of Physics, Jilin University, Changchun, 130012, China}

\author{Daiki Suenaga}
\email{suenaga@hken.phys.nagoya-u.ac.jp}
\affiliation{Department of Physics,  Nagoya University, Nagoya, 464-8602, Japan}

\author{Yusuke Takeda}
\email{takeda@hken.phys.nagoya-u.ac.jp}
\affiliation{Department of Physics,  Nagoya University, Nagoya, 464-8602, Japan}

\date{\today}

\begin{abstract}
We point out that the study of the density dependences of the masses of heavy-light mesons give some clues to the chiral symmetry structure in nuclear matter.
We  include the omega meson effect as well as the sigma meson effect at mean field level on the density dependence of the masses of heavy-light mesons with chiral partner structure. It is found that the omega meson affects the masses of the heavy-light mesons and their antiparticles in the opposite way, while
it affects the masses of chiral partners in the same way.
This is because the $\omega$ meson is sensitive to the baryon number of the light degrees included in the heavy-light mesons.
We also show that
the mass difference between chiral partners is proportional to the mean field of sigma, reflecting the partial restoration of chiral symmetry in the nuclear matter. In addition to the general illustration of the density dependence of the heavy-light meson masses, we
 consider two concrete models for nuclear matter, the parity doublet model and skyrmion crystal model in the sense of mean field approximation.
\end{abstract}



\maketitle

\section{Introduction}

Spontaneous chiral symmetry breaking is one of the most important properties of low energy QCD.
It is expected that the spontaneous chiral symmetry breaking characterized by non-zero value of the quark condensate generates a part of hadron masses and causes the mass splitting between chiral partners.
Then, schematically, hadron masses can be expressed as a sum of the chiral invariant mass and the chiral non-invariant mass coming from the spontaneous chiral symmetry breaking. For example, for the nucleon mass, one
has~\cite{Detar:1988kn,Paeng:2013xya,Ma:2013ooa,Ma:2013ela,Lee:2011tz,Ma:2016gdd}
\begin{eqnarray}
m_N & = & m_0 + \Delta(\langle\bar{q}q\rangle), \nonumber
\end{eqnarray}
where $m_0$ is the chiral invariant mass and $\Delta$ is the part of the mass that vanishes in the chiral symmetric phase. Naturally, it is interesting to ask how much amount of a hadron mass is generated by the chiral symmetry breaking.  An ideal environment to estimate the magnitude of the hadron mass coming from the spontaneous chiral symmetry breaking is QCD at extreme condition in which the chiral symmetry is believed to be partially restored. In such an environment this can be accessed by studying the temperature or/and, which will be done in this work, the density dependence of hadron mass.

In the nucleon sector, by using an effective model with parity doublet structure of baryons, it was found that $\sim 70\%$ of nucleon mass comes from chiral symmetry breaking~\cite{Detar:1988kn}. However, when the baryon as a topological soliton in the hidden local symmetry Lagrangian is immersed in the dense matter which is treated as skyrmion matter, people found that the chiral invariant mass composes $\sim 60\%$ of nucleon mass~\cite{Ma:2013ooa,Ma:2013ela} which roughly the same as that obtained based on the renormalization group analysis of hidden local symmetry Lagrangian with baryons~\cite{Paeng:2013xya}. In this paper, we study the medium modified mass splitting of heavy-light mesons with chiral partner structure in which, it is widely accepted that the mass splitting arises from the spontaneous breaking of chiral symmetry~\cite{Nowak:1992um,Bardeen:1993ae}. Such a kind of study can be tested in the planned experiments in J-PARC, FAIR  and so on.

Studying the properties of heavy-light mesons in medium is also expected to give clues for understanding the chiral symmetry structure (see, e.g., Ref.~\cite{Hosaka:2016ypm} for a review). The medium modified heavy-light meson spectrum has been studied by several groups in the literature~\cite{Harada:2003kt,Mishra:2003se,Friman:2004jh,Yasui:2012rw,%
Suenaga:2014dia,Sasaki:2014asa,Suenaga:2014sga}.
In Ref.~\cite{Suenaga:2014dia}, it was shown that the $D$ meson ($J^P=0^-$) is mixed with $D^\ast$ meson ($J^P=1^-$) in the spin-isospin correlated matter, in which the mixing strength reflects the strength of the correlation. In Refs.~\cite{Harada:2003kt,Sasaki:2014asa,Suenaga:2014sga},
by regarding the $D_0^\ast$ ($J^P=0^+$) and $D_1$ ($J^P=1^+$) mesons as the chiral partners to $D$ and $D^\ast$ mesons, it was shown that the mass splitting of the chiral partner is reduced at high density and temperature.
In particular in Ref.~\cite{Suenaga:2014sga}, by replacing the chiral field for pions interacting with the heavy mesons with its mean field value obtained in the nuclear matter created by the skyrmion crystal approach~\cite{Ma:2013ooa}, it
was shown that the masses of $D$ and $D^\ast$ increase with density while the masses of $D^\ast_0$ and $D_1$ decrease, and that their masses approach the average value. In other word,  the degenerated mass (actually, the difference between the degenerated mass and the heavy quark mass) agrees with the chiral invariant mass,
which is given by the average at vacuum. However, in the analyses of Refs~\cite{Suenaga:2014dia,Suenaga:2014sga}, only the pion is included in the light hadron sector, and effects of other mesons are not included. In particular, the analysis in Ref.~\cite{Mishra:2003se} shows that the $\omega$ meson increases the mass of $D$ meson, while it decreases the $\bar{D}$ meson.

In this paper, we study the effects of $\omega$ meson as well as the $\sigma$ meson on the density dependence of effective masses of heavy-light mesons. We show that the effect of the $\sigma$ meson increases the masses of ($D$, $D^\ast$) heavy quark doublet, while it decreases the masses of the chiral partners, i.e. ($D^\ast_0$, $D_1$) doublet, similarly to the analysis in Refs.~\cite{Sasaki:2014asa,Suenaga:2014sga}.
On the other hand, the effect of the $\omega$ meson increases the masses of both doublets.
Nevertheless, the difference between the masses of chiral partners decreases proportional to the mean field value of the $\sigma$ meson, which reflects the partial chiral symmetry restoration.
As a result, the masses of ($D$, $D^\ast$) doublet and  ($D^\ast_0$, $D_1$) doublet approach a certain degenerate value.  Differently from the previous analysis, the degenerate value does not agree with the average value at vacuum which is the chiral invariant mass of those doublets. In the following analysis, after a general consideration, we consider two concrete models, the parity doublet model~\cite{Motohiro:2015taa} and skyrmion crystal model based on hidden local symmetry~\cite{Ma:2013ooa} to give quantitative results.


\section{Framework}

For explaining the main point explicitly, we work in the heavy quark limit and consider a simple chiral effective model for a heavy meson multiplet of charmed mesons with $J^P = 0^-$, $1^-$, $0^+$ and $1^+$ based on the chiral doubling structure~\cite{Nowak:1992um,Bardeen:1993ae}. Let $H$ and $G$ denote the heavy-quark doublets of heavy-light mesons with the expression
\begin{eqnarray}
H & = & \frac{ 1 + v^\mu \gamma_\mu }{2} \left[ D^{\ast}_{\mu} \gamma^\mu + i D \gamma_5 \right] , \nonumber\\
G & = & \frac{1+v^\mu \gamma_\mu }{2}\left[D_0^\ast - i \gamma^\mu {D}_{1\mu}'\gamma_5 \right] ,
\label{HG doublet}
\end{eqnarray}
where $v^\mu$ is the velocity of the heavy-light mesons, and $D$, $D^{\ast}_{\mu}$, $D_0^\ast$ and $D_{1\mu}^\prime$ are corresponding meson fields. We introduce the chiral fields $\mathcal{H}_{L,R}$ as
\begin{eqnarray}
\mathcal{H}_R = \frac{1}{\sqrt{2}}\left[G + iH\gamma_5\right] \ ,\ \ \mathcal{H}_L = \frac{1}{\sqrt{2}}\left[G - i H\gamma_5\right] ,
\label{eq:HLRGH}
\end{eqnarray}
which transform linearly under the chiral symmetry: ${\mathcal H}_{R,L} \to {\mathcal H}_{R,L} g_{R,L}^\dag$ with $g_{R,L} \in \mbox{SU(2)}_{R,L}$.

The relevant Lagrangian used in the present calculation is expressed as~\cite{Harada:2012km,Suenaga:2014sga}~\footnote{
In Ref.~\cite{Suenaga:2014sga}, another term for the pionic interaction is included. In the present analysis we do not explicitly include the term, since one-pion interaction terms do not contribute to the following analysis.
}
\begin{eqnarray}
{\mathcal L} & = &  {\rm tr}\left[\mathcal{H}_L(iv\cdot\partial)\bar{\mathcal{H}}_L] + {\rm tr}[\mathcal{H}_R(iv\cdot\partial)\bar{\mathcal{H}}_R\right] \notag\\
& & {} - g_{\omega DD} \,\mbox{Tr} \left[ {\mathcal H}_L v^\mu \omega_\mu \bar{\mathcal H}_L +  {\mathcal H}_R v^\mu \omega_\mu \bar{\mathcal H}_R \right] \notag\\
& & {} + \frac{\Delta_M}{2f_\pi} \mbox{tr} \left[\mathcal{H}_L M \bar{\mathcal{H}}_R+\mathcal{H}_R M^{\dagger}\bar{\mathcal{H}}_L\right] \notag\\
& & {} - i\frac{g_{A}}{2f_\pi}{\rm tr}\left[\mathcal{H}_R\gamma_5\gamma^{\mu}\partial_{\mu}M^{\dagger}\bar{\mathcal{H}}_L - \mathcal{H}_L\gamma_5\gamma^{\mu}\partial_{\mu}M\bar{\mathcal{H}}_R\right] ,\nonumber\\
\label{pionlagrangian}
\end{eqnarray}
where $\Delta_M$ is the mass difference between $G$ and $H$ doublets, $f_\pi$ is the pion decay constant, $g_A$ is a dimensionless real parameter.
In the above Lagrangian, the omega meson field $\omega_\mu$ is introduced as a chiral singlet and
the field $M$ is parametrized as $M = \sigma + i \sum_{a=1}^3 \pi_a \tau_a$ with the Pauli matrix $\tau_a$, which transforms as $M \to g_L M g_R^\dag$.
We rewrite the effective Lagrangian~\eqref{pionlagrangian} in terms of $H$ and $G$ fields as
\begin{widetext}
\begin{eqnarray}
{\cal L} & = & {\rm tr}\left[G v^\mu \left(i\partial_\mu + g_{\omega DD} \omega_\mu \right)\bar{G}-H v^\mu \left(i\partial_\mu + g_{\omega DD} \omega_\mu \right) \bar{H}\right]\notag\\
& &{} +\frac{\Delta_M}{4 f_\pi} \mbox{tr} \left[G\left(M+M^{\dagger}\right)\bar{G} + H\left(M+M^{\dagger}\right)\bar{H} - iG\left(M-M^{\dagger}\right)\gamma_5\bar{H} + iH\left(M-M^{\dagger}\right)\gamma_5\bar{G}\right] \notag\\
& &{} - \frac{ig_{A}}{4f_\pi} \mbox{tr} \left[ G\gamma_5\left(\Slash{\partial}M^{\dagger} - \Slash{\partial}M\right)\bar{G} -H\gamma_5\left(\Slash{\partial}M^{\dagger}-\Slash{\partial}M\right)\bar{H} + iG\left(\Slash{\partial}M^{\dagger}+\Slash{\partial}M\right)\bar{H}-iH\left(\Slash{\partial}M^{\dagger}+\Slash{\partial}M\right)\bar{G} \right]
\ . \label{pionlagrangian2}
\end{eqnarray}
\end{widetext}

Now, we replace the light meson fields by their mean field values in medium.
Here we consider the symmetric matter only and assume no pion condensation, so that
$\langle M \rangle = \langle \sigma \rangle$ and $\langle \partial_\mu M \rangle = 0$.
Note that the mean field value of $\sigma$ at vacuum agrees with the pion decay constant, $\langle \sigma \rangle_0 = f_\pi$. From the above form, we obtain
\begin{eqnarray}
{\cal L}_{\rm eff} & = & {\rm tr}\left[G \left(i\partial_0 + g_{\omega DD} \langle \omega_0 \rangle \right) \bar{G} \right] \notag\\
& &{} - {\rm tr}\left[ H \left(i\partial_0 + g_{\omega DD} \langle \omega_0 \rangle \right)  \bar{H} \right] \notag\\
& &{} + \frac{\Delta_M}{2f_\pi} \left\langle \sigma \right\rangle \mbox{tr} \left[G\bar{G}+H\bar{H}\right] , \label{eq:ga1ga20}
\end{eqnarray}
where we used $v^\mu = ( 1, \vec{0})$. So that, the effective masses of $H$ and $G$ doublets are obtained as
\begin{eqnarray}
m_H^{\rm(eff)} & = & m - \frac{\Delta_M}{2f_\pi}  \langle \sigma \rangle + g_{\omega DD} \langle \omega_0 \rangle \ , \nonumber\\
m_G^{\rm(eff)} & = & m + \frac{\Delta_M}{2f_\pi}  \langle \sigma \rangle + g_{\omega DD} \langle \omega_0 \rangle \ ,
\label{eff mass}
\end{eqnarray}
where $m$ is the average mass of the $H$ and $G$ doublets with $m = (m_H + m_G)/2$. The masses of $H$ and $G$ doublets are determined by the spin average of the physical masses as
\begin{eqnarray}
m_H = \frac{m_D + 3 m_{D^\ast} }{4} \ , \quad
m_G  = \frac{m_{D_0^\ast} + 3m_{D_1}}{4}\ .
\end{eqnarray}
We should note that, for the anti-charmed mesons $\bar{D}$, $\bar{D}^\ast$, $\bar{D}_0^\ast$ and $\bar{D}_1$, the sign in front of the coupling to the omega meson is flipped, so that the effective masses are written as
\begin{eqnarray}
m_{\bar{H}}^{\rm(eff)} & = & m - \frac{\Delta_M}{2f_\pi}  \langle \sigma \rangle - g_{\omega DD} \langle \omega_0 \rangle \ , \nonumber\\
m_{\bar{G}}^{\rm(eff)} & = & m + \frac{\Delta_M}{2f_\pi}  \langle \sigma \rangle - g_{\omega DD} \langle \omega_0 \rangle \ .
\label{eff mass bar}
\end{eqnarray}

Now, let us study the density dependence of masses in Eqs.~(\ref{eff mass}) and (\ref{eff mass bar}).
As for the mean field value of $\omega$, we simply take
\begin{equation}
\langle \omega_0 \rangle = \frac{g_{\omega NN} }{ m_\omega^2 } \rho_B \ ,
\label{omega rhoB}
\end{equation}
where $g_{\omega NN}$ is the omega meson coupling to the nucleon, $m_\omega$ is the mass of omega meson and $\rho_B$ is the baryon number density.
As for the mean field of $\sigma$ we adopt the linear densty approximation as
\begin{equation}
\frac{\langle \sigma \rangle}{ \langle \sigma \rangle_0} = 1 - \frac{\sigma_{\pi N}}{ m_\pi^2 f_\pi^2 } \rho_B \ ,
\end{equation}
where $\sigma_{\pi N}$ is the coefficient of the $\pi$-$N$ sigma term.

For making a numerical estimation, we use $m_G=2.40\,$GeV, $m_H = 1.97$~GeV, and $\Delta_M= m_G - m_H = 430~\mbox{MeV}$ for masses, in addition to $m_\omega = 783$~MeV,  $m_\pi = 137$\,MeV and $f_\pi =92.1\,$MeV. As for other parameters, we use $\sigma_{\pi N} = 45\,$MeV, $\vert g_{\omega DD} \vert = 3.7$ estimated in Appendix~\ref{app:A} and $\vert g_{\omega NN} \vert = 6.23$,~\footnote{
There are several values listed in literatures.  Here, we use a value obtained in an analysis of nuclear matter based on the parity doublet model in Ref.~\cite{Motohiro:2015taa}, in which the saturation density, the binding energy and the incompressibility are reproduced.
Table~I in the paper includes some errors, and the value $\vert g_{\omega NN} \vert = 6.23$ is the corrected value obtained for the chiral invariant mass $m_0 = 700$\,MeV.
}
which lead to $\vert g_{\omega NN} g_{\omega DD} \vert = 23 $,
as reference values.
We note that the $D$ meson includes the anti-light quark, and $\bar{D}$ meson does the light quark.
Therefore, it is natural to consider that the $\bar{D}$ meson is affected by the Pauli blocking in a dense medium, which is represented by the effect of the mean field of the $\omega$ meson.
Then, for
 the concreteness of the discussion, we take $g_{\omega NN} g_{\omega DD} < 0 $ below.
When we study the case with $g_{\omega NN} g_{\omega DD} > 0 $, we just exchange $H$ with $\bar{H}$ and $G$ with $\bar{G}$ in the following discussion.

\begin{figure}[htbp]
\includegraphics[width = 7cm]{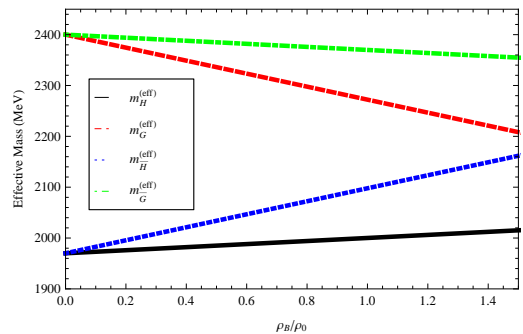}
\caption{(Color online) Density dependence of the effective masses of
$H$ doublet (black curve), $\bar{H}$ doublet (blue curve), $G$ doublet (red curve) and $\bar{G}$ doublet (green curve) with $\sigma_{\pi N} = 45\,$MeV and $g_{\omega DD} g_{\omega NN} =- 23$.}
 \label{fig:mass}
\end{figure}

We plot the density dependence of the masses in Fig.~\ref{fig:mass}.
This shows that the masses of $H$ and $\bar{H}$ doublets as well as those of $G$ and $\bar{G}$ doublets are split by $\omega$ contribution.

\begin{figure}[htbp]
\includegraphics[width = 8cm]{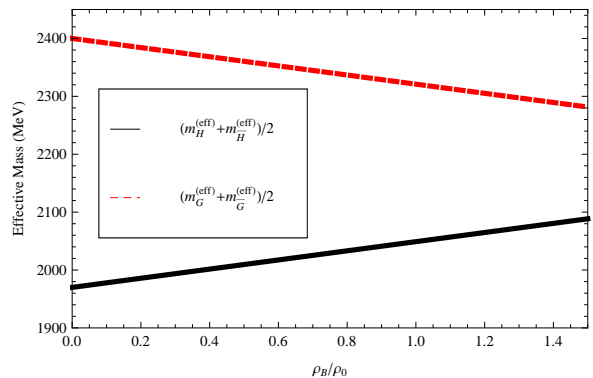}
\caption{(Color online) Density dependence of the effective masses of charmed mesons.
The red-dashed and black-solid curves show the sums of $m_{G}^{\rm(eff)} + \overline{m}_{G}^{\rm(eff)} $ and $m_{H}^{\rm(eff)} + \overline{m}_{H}^{\rm(eff)}$ divided by two, respectively, which do not depend on the sign of $g_{\omega DD} g_{\omega NN}$.
}
 \label{fig:sum of mass}
\end{figure}

From the $\omega$ contribution combined with the $\sigma$ contribution,
the mass of $G$ doublet (indicated by red-dashed curve) decreases with increasing density, and the $\bar{H}$ mass (by blue-dotted curve) increases.
On the other hand, $H$ mass (by black-solid curve) and $\bar{G}$ mass (by green-dotdashed curve) are rather stable.
Aa a result, the $G$ mass tends to degenerate with the mass of $H$ doublet at certain high density.
If one measures the mass of $H$ only, one might think that the chiral invariant mass would be almost same as the mass of $H$ doublet. However, the actual chiral invariant mass is larger than the $H$ mass at vacuum, which can be obtained by averaging the masses of the particles ($G$ and $H$) and anti-particles ($\bar{H}$ and $\bar{G}$), as shown in Fig.~\ref{fig:sum of mass}.
We should note that the sums of masses of particle and anti-particle are actually independent of the sign of    $g_{\omega DD} g_{\omega NN}$.

\begin{figure}[htbp]
\includegraphics[width = 6cm]{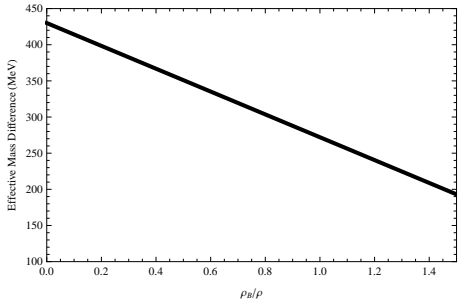}
\caption{Density dependence of the difference of the effective masses of charmed mesons defined by Eq.~(\ref{MassDifference}).}
 \label{fig:diff}
\end{figure}

Mass difference between the chiral partners, i.e., the $H$ doublet and the $G$ doublet, is caused by the spontaneous chiral symmetry breaking. This structure is seen by subtracting the mass of $H$ doublet from that of $G$ doublet with Eq.~(\ref{eff mass}) as
\begin{eqnarray}
m^{({\rm eff})}_G-m_H^{({\rm eff})} = \frac{\Delta_M}{f_\pi}\langle\sigma\rangle\ . \label{MassDifference}
\end{eqnarray}
So the mass difference is expected to give a clue for the chiral condensate.  In the mean field approximation, it is actually proportional to the mean field $\langle \sigma \rangle$ as shown in Fig.~\ref{fig:diff}. This figure clearly shows that, with the increasing of the nuclear matter density, chiral symmetry is (partially) restored.

\begin{figure}[htbp]
\begin{center}
\includegraphics[width = 7cm]{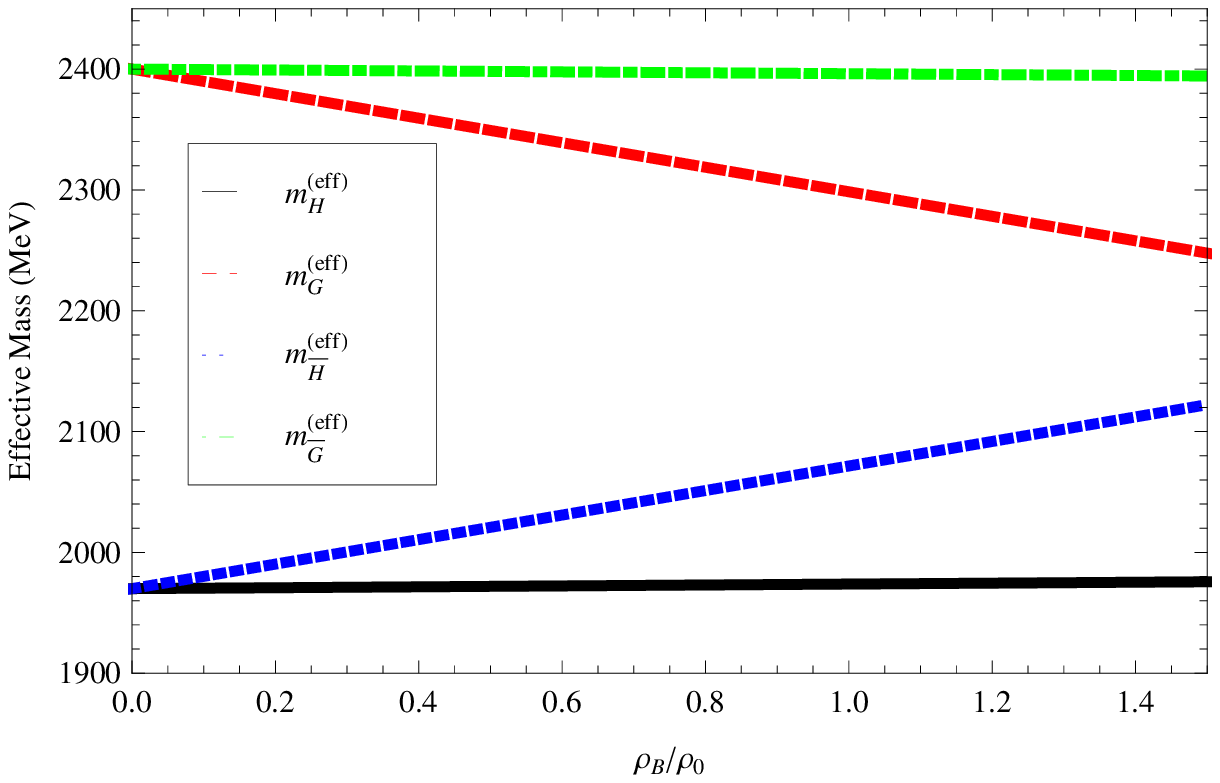}\qquad\quad
\includegraphics[width = 7cm]{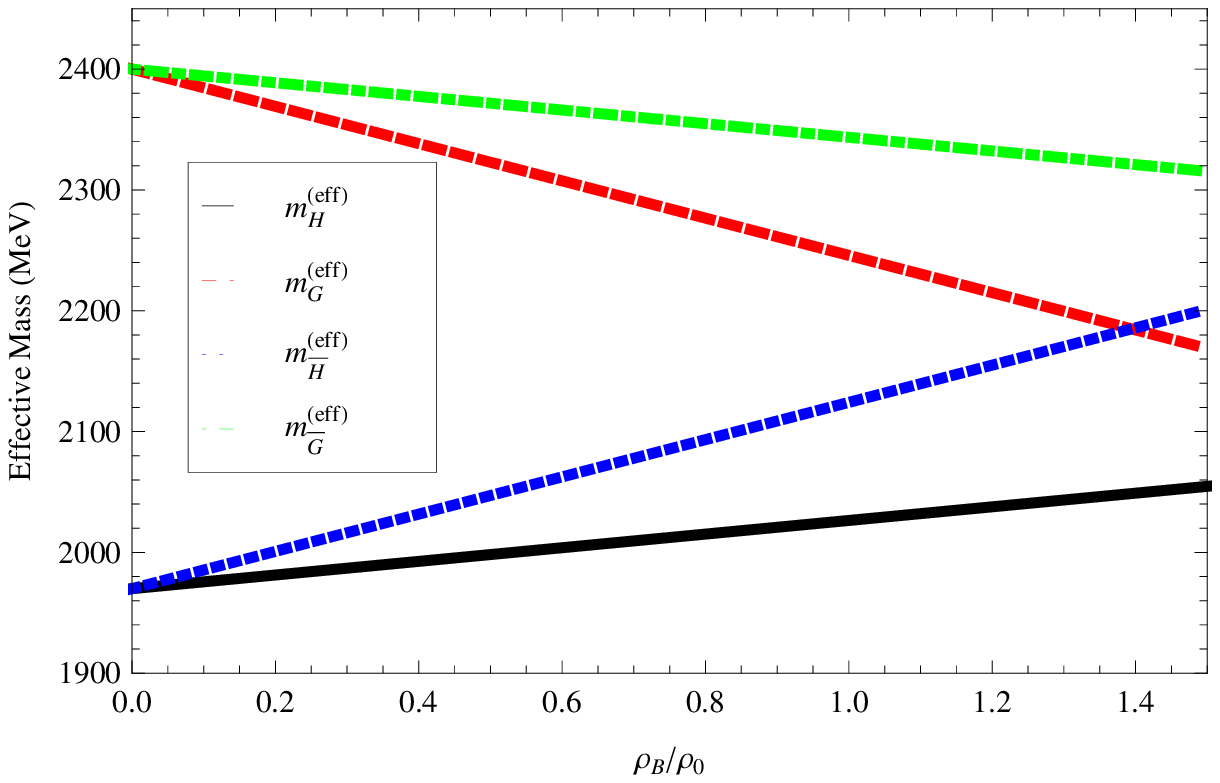}\\
\end{center}
\caption{(Color online) Density dependence of the effective masses of charmed mesons for
with $\sigma_{\pi N} = 30\,$MeV (upper panel) and $\sigma_{\pi N} = 60\,$MeV (lower panel). Notations are the same as in Fig.~\ref{fig:mass}.
}
 \label{fig:mass piN}
\end{figure}

For checking the $\pi$-$N$ sigma term dependence of the effective masses, we vary the value of $\sigma_{\pi N}$ as $30$ and $60$\,MeV, which are plotted in Fig.~\ref{fig:mass piN}.
This shows that the difference between the masses of $H$ and $G$ as well as that between $\bar{H}$ and $\bar{G}$ decreases more rapidly for larger value of $\sigma_{\pi N}$.  As a result, the chiral symmetry restores more rapidly for the larger $\sigma_{\pi N}$.

\begin{figure}[htbp]
\begin{center}
\includegraphics[width = 7cm]{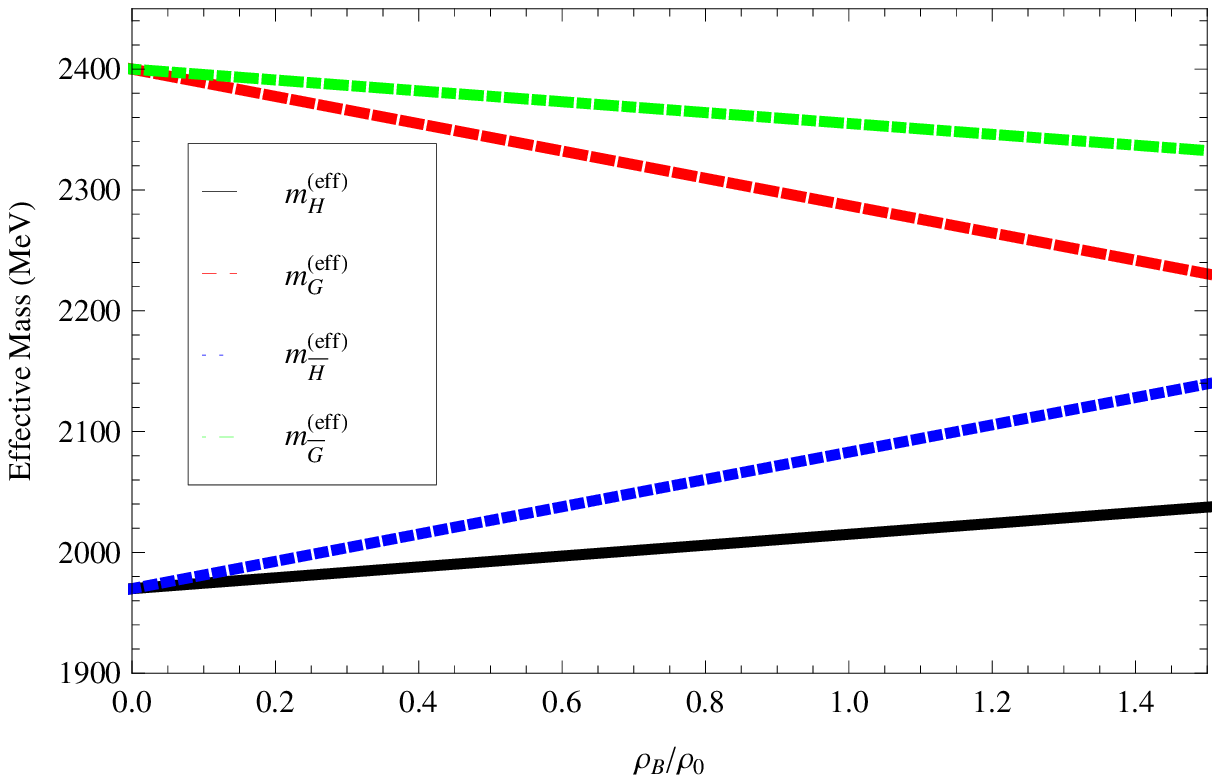}\qquad\quad
\includegraphics[width = 7cm]{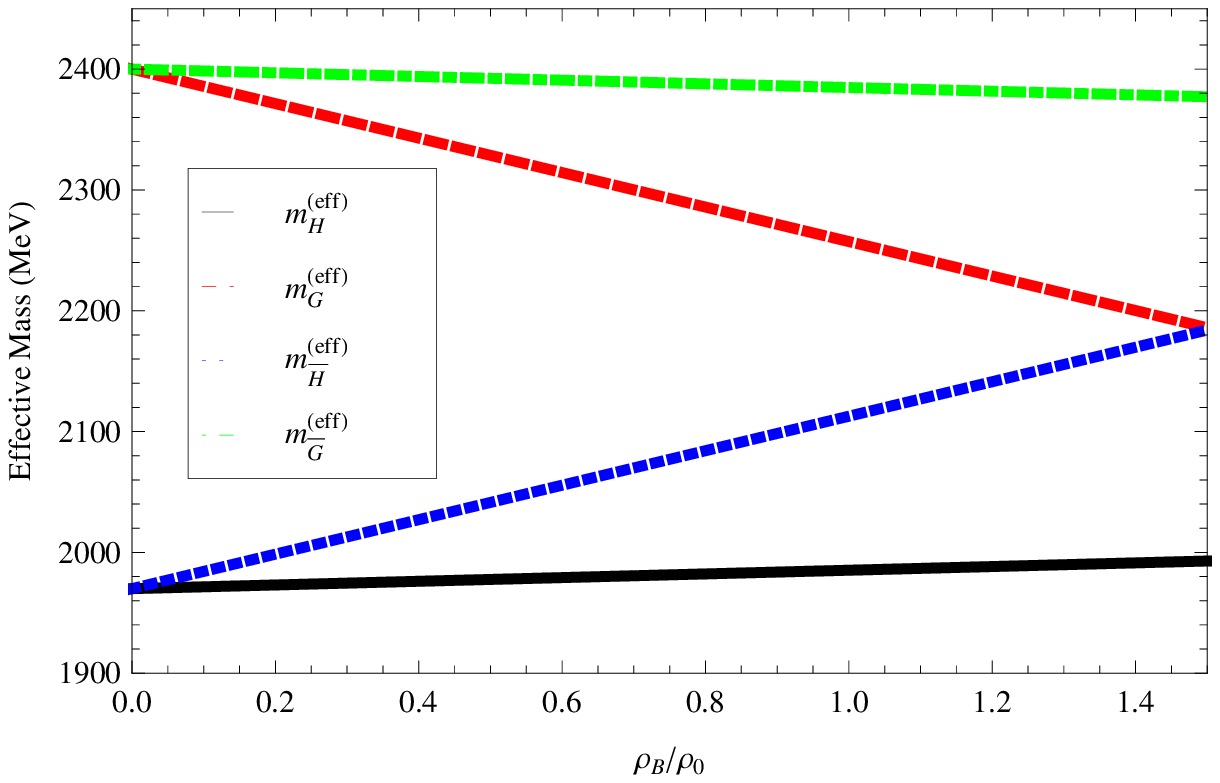}\\
\end{center}
\caption{(Color online) Density dependence of the effective masses of charmed mesons for $\sigma_{\pi N} = 45\,$MeV with $ g_{\omega DD} g_{\omega NN} = - 16$ (upper panel), and $g_{\omega DD} g_{\omega NN} = - 30$ (lower panel). Notations are the same as in Fig.~\ref{fig:mass}.
}
 \label{fig:mass gg}
\end{figure}

We next check the dependence on the value of $\vert g_{\omega DD} g_{\omega NN}\vert$ in Fig.~\ref{fig:mass gg}, by taking 30\% deviation from the estimated value.
This shows that the masses change more rapidly for larger value of  $\vert g_{\omega DD} g_{\omega NN}\vert$.

\section{Model analysis}

After the above general discussion, let us study the density dependences of the effective masses based on some specific models. Here we use the the nuclear matter described by the parity doublet model~\cite{Motohiro:2015taa} and by the skyrmion crystal model based on the hidden local symmetry~\cite{Ma:2013ooa}.

\subsection{Parity doublet model}

In Ref.~\cite{Motohiro:2015taa},  the parity doublet model based on the linear $\sigma$ model~\cite{Detar:1988kn}, in which an excited nuclear with negative parity, $N^\ast(1535)$, is regarded as the chiral partner to the ordinary nucleon, was extended by including a six-point interaction for $\sigma$ field and interactions to the $\omega$ and $\rho$ mesons based on the hidden local symmetry, to study the nuclear matter.
It was shown that, for wide range of the chiral invariant mass for the nucleon, the model reproduces the saturation density, binding energy, incompressibility and  symmetry energy.
In Refs.~\cite{Motohiro:thesis} and \cite{TKH}, it is shown that the ratio of the mean field $\langle \sigma \rangle$ at normal nuclear matter density to the one at vacuum obtained for the chiral invariant mass of nucleon $m_{0}=500$\,MeV is consistent with the experimental value of the one for the pion decay constant~\cite{Suzuki:2002ae,Kienle:2004hq}.
Here we use the density dependences of $\langle \sigma \rangle$ and $\langle  \omega \rangle$ obtained from the model with $m_{0} = 500$\,MeV.

\begin{figure}[htbp]
\begin{center}
\includegraphics[width = 7cm]{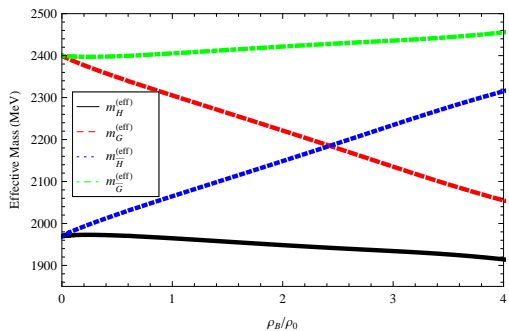}
\end{center}
\caption{(Color online) Density dependence of the effective masses of charmed mesons in parity doublet model. Notations are the same as in Fig.~\ref{fig:mass}.
}
 \label{fig:masspdm}
\end{figure}

\begin{figure}[htbp]
\includegraphics[width = 7cm]{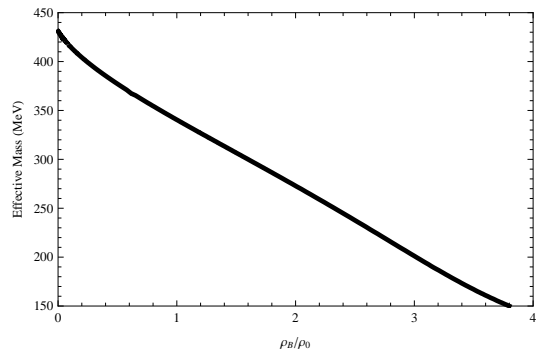}
\caption{
Density dependence of the difference of the effective masses of charmed mesons defined by Eq.~(\ref{MassDifference}) in the parity doublet model.
}
 \label{fig:diff PDM}
\end{figure}

We show the resultant density dependence of masses in Fig.~\ref{fig:masspdm}.
Here we use $\vert g_{\omega DD} \vert = 3.7$ estimated in Appendix~\ref{app:A} as a typical value.
This shows that
the density dependence of all masses in the very low density region
 $\rho_B/\rho_0 \lesssim 0.3$
is similar to the one in Fig.~\ref{fig:mass} reflecting that both $\langle \sigma \rangle$ and $\langle \omega \rangle$ in the parity doublet model are consistent with those obtained in the linear density approximation
as can be seen in Ref.~\cite{Motohiro:thesis}.
However,
around the density region $\rho_B /\rho_0 \sim 0.3$,
the mass of $H$ doublet (black curve) starts to decrease and that of $G$ doublet (green curve) to increase, differently from the linear density approximation.
In this model, the mean field $\langle \omega \rangle$ is proportional to the density similarly to the linear density approximation in Eq.~(\ref{omega rhoB}).  Then, the different density dependence of the masses is originated in $\langle \sigma \rangle$.
In Fig.~\ref{fig:diff PDM}, we plot
the difference of two masses of $H$ and $G$ doublets.
This shows that the difference decreases more slowly than that in the linear approximation shown in
Fig.~\ref{fig:diff}.

\subsection{Skyrme model}

In Refs.~\cite{Ma:2013ooa,Ma:2013ela}, the skyrmion crystal model is used to study the qualitative structure of  nuclear matter
by regarding the skymrion matter as nuclear matter in the sense of large $N_c$ limit of QCD. A robust conclusion drown in the skyrmion crystal approach is that, when the density of the nuclear matter is increased, the skyrmion matter undergoes a topological phase transition to the matter made of half-skyrmions in which the space average of the chiral condensate vanishes although it is locally non-zero and the chiral symmetry is still broken~\cite{Harada:2015lma}.
 Since the half-skyrmion phase is not observed in nature, the critical density should be higher than the normal nuclear density. Recently, the description of nuclear matter from the skyrmion crystal and the implication of the topological phase transition in the equation of state of neutron star have gotten great progress (see, e.g., Ref.~\cite{Ma:2016gdd} for a recent review).

\begin{figure}[htbp]
\begin{center}
\includegraphics[width = 7cm]{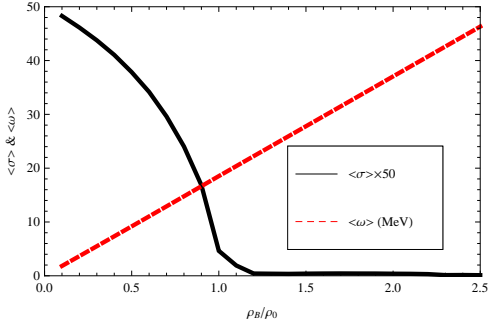}
\end{center}
\caption{(Color online) Density dependence of $\langle \sigma \rangle$  and $\langle \omega \rangle$ calculated in skyrmion crystal model.
}
 \label{fig:vevSW}
\end{figure}

In the present analysis, we calculate the mean fields $\langle \sigma \rangle$ and $\langle \omega \rangle$ in the skyrmion crystal model, and substitute the values into Eqs.~(\ref{eff mass}) and (\ref{eff mass bar}) to obtain the density dependence of the charmed meson masses.
\begin{figure}[htbp]
\begin{center}
\includegraphics[width = 7cm]{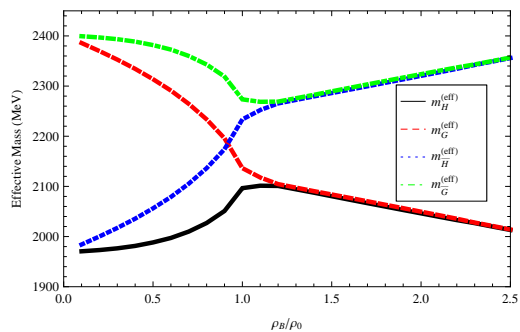}
\end{center}
\caption{(Color online) Density dependence of the effective mass of charmed mesons in skyrmion crystal model with $g_{\omega DD} ={} - 3.7$. Notations are the same as in Fig.~\ref{fig:mass}.
}
 \label{fig:massskyr}
\end{figure}
We plot in Fig.~\ref{fig:massskyr} the density dependence of the effective masses of charmed mesons by using $\langle \sigma\rangle$ and $\langle \omega\rangle$ with $g_{\omega DD} ={} - 3.7$ calculated by the skyrmion crystal model based on the hidden local symmetry~\cite{Ma:2013ooa} and other parameters are the same as that used in the plot of Fig.~\ref{fig:mass}. In this model, we find that both $G$ and $\bar{G}$ masses decrease with density while both $H$ and $\bar{H}$ increase with density. Because the density dependence of $\bar{H}$ mass and $G$ mass is stronger than that of $\bar{G}$ mass and $H$ mass, $H$ and $G$ as well as  $\bar{H}$ and $\bar{G}$ become degenerate at density $\sim 1.2 \rho_0$ at which the skyrmion phase transits to half-skyrmion phase. This is because, the mass difference between $H$ and $G$ as well as  $\bar{H}$ and $\bar{G}$ is proportional to $\langle \sigma \rangle$ which vanishes in the half-skymrion phase. Moreover, we find that the degenerated mass of $H$ and $G$ and that of $\bar{H}$ and $\bar{G}$ linearly depend on density in the half-skyrmion phase. The reason is that, the $\langle \omega \rangle$ is a linear function of density and this linear dependence agrees Eq.~\eqref{omega rhoB}. We plot in Fig.~\ref{fig:vevSW} the density dependence of $\langle \sigma \rangle$ and $\langle \omega \rangle$. 
Note that, as stressed above,
 the analysis shows just a qualitative structure,
 and the degeneracy of chiral partners does not imply the chiral restoraton but due to the vanish of the space average of chiral condensate in the half-skyrmion matter.
 
\section{A summary and discussion}

In this work, by regarding the ($D^\ast_0$, $D_1$) heavy quark doublet as the chiral partner of the  ($D$, $D^\ast$) doublet, we explicitly showed that the effect of the $\omega$ meson decreases
the masses of both doublet, while ($\bar{D}^\ast_0$, $\bar{D}_1$) and ($\bar{D}$, $\bar{D}^\ast$) meson masses are increased.
We explicitly point out that the $\omega$ meson effect is significant for understanding the density dependence of  effective hadron masses in medium. Even though the qualitative dependence is model dependent, the tendency that the masses of the heavy-light mesons  and their antiparticles are  split  due to the $\omega$ meson effect is robust. We hope this medium modified masses of the heavy-light mesons can be detected in the future experiments at J-PARC and FAIR through the strong and weak channels, such as $\psi(3770) \to D\bar{D}$, $J/\psi \to \bar{D} e^+ \nu_e$ and so on. (See, e.g., Ref.~\cite{Ohnishi}.)

We would like to note that the result of the omega meson effect to $D$ and $\bar{D}$ mesons are consistent with the result obtained in Ref.~\cite{Mishra:2003se}. In our analysis, we further
introduce $p$-wave excited $D$ and $\bar{D}$ mesons by using the chiral doubling model, and we found that the difference between the masses of chiral partners decreases in proportion to the mean field value of the $\sigma$ meson, which reflects the partial chiral symmetry restoration even if the $\omega$ meson contribution enters.

In our calculation, we simply take the mean field approach.  An extension of the present work to include some loop contributions will be reported in~\cite{SYH}. In the present work, we only discussed the medium modified charmed mesons. The results presented here are intact for their bottom cousins except the average mass $m$ should be taken the value of bottom mesons.

\appendix

\section{Estimation of $g_{\omega DD}$}
\label{app:A}

In the present analysis, we estimate a reference value of $g_{{\omega DD}}$ defined by Eq.~\eqref{pionlagrangian} in the heavy hadron limit by using the following naive scaling property:
\begin{equation}
\left\vert \frac{ \tilde{g}_{{\omega D\bar{D}}} }{ \tilde{g}_{\omega K\bar{K}} } \right\vert = \left\vert \frac{ \tilde{g}_{{D^{\ast+} D^0 \pi^-}} }{ \tilde{g}_{{K^{\ast0} K^- \pi^+}} } \right\vert
\ ,
\label{VPP Lagrangian}
\end{equation}
where the coupling constants are defined in the relativistic form of the interaction Lagrangian among a vector meson $V$ and two pseudoscalar mesons $P$ and $P'$ espressed as
\begin{equation}
{\mathcal L}_{VPP'} = i \tilde{g}_{{V P P'}} V^\mu \left( \partial_\mu P P' - \partial_\mu P' P \right) \ .
\end{equation}
The coupling $\tilde{g}_{D^{\ast+} D^0 \pi^+}$ appears in the decay width of $D^{\ast +} \to D^0 \pi^+$ as
\begin{equation}
\Gamma( D^{\ast +} \to D^0\pi^+) = \frac{ \tilde{g}_{D^*D\pi}^2|\vec{p}|^3}{24\pi m_H^2}=56.5 \ {\rm keV}\ ,
\end{equation}
which with $|\vec{p}|=39.4$\,MeV and $m_H = 1.97$\,GeV leads to
\begin{eqnarray}
|\tilde{g}_{D^*D\pi}| = 16.5\ .\label{CouplingDstDPi}
\end{eqnarray}
In a class of three-flavor chiral models for vector mesons $g_{\omega K\bar{K}}$ and $g_{K^{\ast0} K^- \pi^+}$ are related to the vector meson masses as~\cite{Harada:1995sj}
\begin{align}
& g_{\omega K^+ K^-} = g_{\omega K^0 \bar{K}^0 } = g_{\omega K \bar{K}} = \frac{1}{4} \frac{m_\omega^2}{g f_K^2} \ , \notag\\
& g_{K^{\ast0} K^- \pi^+} = \frac{1}{2\sqrt{2}} \frac{ m_{K^\ast}^2 }{ g f_K f_\pi} \ ,
\label{eq:K couplings}
\end{align}
where $m_\omega$ and $m_{K^\ast}$ are the masses of $\omega$ and $K^\ast$ mesons, $f_\pi$ and $f_K$ are the pion and kaon decay constants, and $g$ is the gauge coupling constant of the hidden local symmetry~\cite{Harada:2003jx}.
Using
$m_\omega = 783$\,MeV,
$m_{K^{\ast0}} = 896$\,MeV,
$f_\pi = 92.1$\,MeV and
$f_K = 110$\,MeV,
the ratio of two couplings in Eq.~(\ref{eq:K couplings}) is estimated as
\begin{equation}
\left\vert \frac{ g_{\omega K\bar{K}} }{ g_{K^{\ast0}K^- \pi^+} } \right\vert = 0.452 \ ,
\end{equation}
which with Eq.~(\ref{CouplingDstDPi}) leads to
\begin{equation}
\left\vert \tilde{g}_{{\omega D\bar{D}}} \right\vert = \left\vert \frac{ \tilde{g}_{\omega K\bar{K}} }{ \tilde{g}_{{K^{\ast0} K^- \pi^+}} } \tilde{g}_{{D^{\ast+} D^0 \pi^-}}
\right\vert = 7.4
\ ,
\end{equation}

From the heavy quark Lagrangian in Eq.~(\ref{pionlagrangian2}), the $\omega$-$D$-$\bar{D}$ interaction is written as
\begin{equation}
{\cal L}_{\omega DD} =  2 g_{\omega DD} D \, \omega_\mu v^\mu \bar{D} \ ,
\end{equation}
with a scaling factor of the mass of heavy meson $M_H$.
Comparing this with Eq.~(\ref{VPP Lagrangian}), we estimate $g_{\omega DD}$ as
\begin{equation}
\left \vert g_{\omega DD} \right\vert = \frac{1}{2} \left\vert \tilde{g}_{{\omega D\bar{D}}} \right\vert = 3.7 \ ,
\end{equation}
which is the value used in the present work.

\acknowledgments

 The work of M.~H. was supported in part by the JSPS Grant-in-Aid for Scientific Research (C) No.~16K05345. Y.-L.~M. was supported in part by National Science Foundation of China (NSFC) under Grant No. 11475071, 11547308 and the Seeds Funding of Jilin University.


\end{document}